\documentclass[a4paper,aps,twocolumn,nofootinbib,nobibnotes,twoside]{revtex4-1}

\usepackage{mathrsfs}

\usepackage{array}
\usepackage{geometry}
\usepackage{amsmath}
\usepackage{amsthm}
\usepackage{bm}
\usepackage{bbm} 
\usepackage{amssymb}
\usepackage{xspace}
\usepackage{amscd}
\usepackage{dsfont}
\usepackage{t1enc}
\usepackage{scalefnt}
\usepackage[greek, english]{babel}
\usepackage[latin2]{inputenc}
\usepackage{teubner}
\usepackage{graphicx}
\usepackage{multirow}
\usepackage{dcolumn}
\usepackage{braket}
\usepackage{xifthen}
\usepackage{textcomp}
\usepackage{tikz}
\usepackage{tipa}
\usepackage{pifont}
\usepackage{cancel}
\usepackage{stackrel}
\usepackage{xcolor}
\usepackage{soul}
\usepackage{stmaryrd}
\usepackage{multibib}

\usetikzlibrary{matrix,arrows,decorations.markings,decorations.pathreplacing,petri,topaths}
\usepackage[stable,symbol]{footmisc}
\definecolor{col}{rgb}{0.25, 0.41, 0.88}
\usepackage[pdftex, pdftitle={}, pdfauthor={}, colorlinks=true, urlcolor=col, linkcolor=col, bookmarks=false, citecolor=col, raiselinks=false,hyperfootnotes=false]{hyperref}

\usepackage{actaneasy}
\usepackage[T1]{fontenc}

\begin{document}
\vs*{50pt}

\ch{Electron Localization in Rydberg States}
\aum{J. Mostowski et al.}
\arm{Electron Localization in Rydberg States}
\spt{J. Mostowski, J. Pietraszewicz}{Electron Localization in Rydberg States}
\aut{J. Mostowski$^*$ and J. Pietraszewicz}{Institute of Physics, Polish Academy of Sciences, al. Lotnik\'ow 32/46, PL-02668 Warsaw, Poland}

\pacc{We discuss the possibility of localizing an~electron in a~highly excited Rydberg state. The second-order correlation of emitted photons is the tool for the determination of electron position. This second-order correlation of emitted radiation and, therefore, the correlation of operators describing the acceleration of the electron allows for a~partial localization of the electron in its orbit. The correlation function is found by approximating the transition matrix elements by their values in the classical limit. It is shown that the second-order correlation, depending on two times, is a~function of the time difference and is a~periodic function of this argument with the period equal to the period of the corresponding classical motion. The function has sharp maxima corresponding to large electron acceleration in the vicinity of the ``perihelion.'' This allows the localization of the electron in its consecutive approach to the perihelion point.}{Rydberg state, radiation, second order correlation, localization\vs*{55pt}}{jan.mostowski@ifpan.edu.pl}

\section{Introduction\vs*{6pt}}
 
The measurement process, since the early days of quantum physics, has been one of the central issues in many attempts to understand the relation between the classical and quantum description of physical systems~[1] (for a~more recent analysis, see, e.g.,~[2]). The most common theory of quantum measurement~\m{[3--5]}~assumes that the quantum system is coupled to a~meter. The interaction between them entangles the two systems. The measurement, described as a~projection onto the state of the meter, provides information on the state of the system. Continuous measurements of quantum systems treated as stochastic processes were first considered in~[6] in the context of photon counting. The formalism based on path integrals was initiated in~[7] and further developed in~[4] (see also~[8]).

The classical motion of the electron bound in a~Coulomb field is periodic. The wavefunction describing the bound electron in a~stationary state does not show any time-dependent features. Time dependence, and hence classical features of wavefunctions, can be obtained for non-stationary states, linear combinations of energy eigenstates with different energies. Such a~construction is well known in the case of a~harmonic oscillator, and the most \m{classical} states are well-known coherent states~[9] (see also, e.g.,~[10]). The corresponding time-dependent states in the case of Rydberg states were introduced in~[11] (see also~\m{[12--15]}).

Another point of view was presented in~[16], where it is pointed out that when a~measurement breaks the time-translational symmetry of a~stationary state, a~periodic motion of the system is initiated. This approach was further elaborated in~[17,~18].

The classical limit of quantum mechanics is still a~vivid subject of investigation (see, e.g.,~[19]). One of the recently discussed problems in this area relates to the successive measurements of particle position and detection of the trajectory. Most of the interest has been limited to free particles, and not much has been done in the case of bound states.

Quantum description of the hydrogen atom is well known. All energies and wavefunctions of stationary states are well known. The classical limit is approached in the limit of large quantum numbers --- the wavefunction should be related to classical trajectories. This relation has been discussed in many papers. Both time-dependent states, analogs of harmonic oscillator coherent states, and stationary states in the limit of high excitation were shown to exhibit classical features.

In this paper, we will present yet another aspect of the classical limit in the case of Rydberg states. Namely, we will use the radiation emitted from the highly exited state to determine the electron position as a~function of time. Detection of radiation at a~given time breaks the time-translational symmetry and allows observation of the time dependence of subsequent evolution. This approach provides a~partial but straightforward way of estimating the elements of the time-dependent classical trajectory hidden in the stationary wavefunction.

Radiation from a~quantum system, such as a~hydrogen atom, is usually studied in the frequency domain. The spectrum consists of several lines. Measurement of the spectrum is not the only possibility --- time dependence of radiation can be studied as well. The time dependence of the spontaneous emission from a~highly excited vibrational state of a~diatomic molecule was used to \m{determine} the time-dependent relative position of the constituents. This allowed us to demonstrate the time dependence of various states, such as coherent states and others, e.g., the Schr\"odinger cat state~[20]. Let us note that in the case of Rydberg states with the principle quantum number $n\approx100$, the characteristic frequency of radiation is $\nu\approx 10^{10}$~Hz, so the time dependence of the radiation for times smaller than $1/\nu$ is within experimental reach. Radiation observed for such small times of the order of $1/\nu$ exhibits different features as compared to the long-time measurements. This and the relation to the position measurement will be discussed below.
\vs*{4pt}

\section{A simple case --- harmonic oscillator\vs*{4pt}}

We will begin the discussion of electromagnetic radiation in the time domain and its relation to the measurement of the electron position with a~simple example of a~harmonic oscillator. The charged particle oscillates with the frequency $\omega$ along the $x$ axis; its motion is given by $x_{cl}(t)=A\cos(\omega t)$. This electron is a~source of electromagnetic radiation. We~will find the $x$ component of the electric field in the far zone along the $y$ axis (to simplify the geometry). We~have, in the dipole approximation,
\beq 
E_x(\we{R},t) = -\frac{e}{4\pi\epsilon_0 R}\ \omega^2 x_{cl}(t_{ret}),
\eeq{1}
where $t_{ret}=t-|\we{R}|/{\rm c}$ is the retarded time, and $c$ is the speed of light. We have skipped the $R$ dependence of the field --- it is just like in classical electrodynamics, namely $E_x \sim R^{-1}$. It follows from~(1) that the electric field oscillates with the frequency~$\omega$. This classical treatment does not take into account radiation damping, thus it is valid only for a~short time, shorter than the characteristic damping time.

We will now discuss an~emission of radiation, taking into account the quantum nature of the oscillator. We will concentrate on the highly excited states of the oscillator and hence on the classical limit.

The position of an~oscillating particle is described by the position operator $x$. It can be expressed in terms of the lowering and raising operators $a$ and $a^{\dagger}$, respectively, as follows
\beq 
x=x_0\, \frac{(a+a^{\dagger})}{\sqrt{2}},
\eeq{2}\\[3pt]
where $x_0=\sqrt{\frac{\hbar}{M\omega}}$, $\hbar$ is the Planck constant, and $M$ denotes the mass of the oscillating particle. The component $E_x$ of the electric field operator (the radiated part) in the dipole approximation is given by
\beq 
E_x(\we{R},t)= -\frac{e}{4\pi\epsilon_0 R}\ \omega^2 x(t_{ret}),
\eeq{3}\\[3pt]
just like in the classical case. This time, however, the electric field is an~operator, and we will find the expectation values of this operator. We assume that at time $t=0$, the oscillator is in the energy eigenstate $|n\rangle$ with energy $E_n=\hbar\omega n$. Thus the expectation value of the $x$ operator, and hence of the $E_x(\we{r},t)$ operator, is equal to zero. The first-order correlation function becomes
\beq 
\big\langle E_x(\we{R},t_2) E_x(\we{R},t_1)\big\rangle =\frac{1}{2}\,\frac{e^2}{(4\pi\epsilon_0 R)^2}\, \omega^4x_0^2
\eeq{}\tylx
\beq\quad \times \Big[n \, \ee^{\ii\omega (t_2{-}t_1)}+\Big(n+\frac{1}{2}\Big)\, \ee^{-\ii\omega (t_2{-}t_1)} \Big].
\eeq{4}\tylx

In the case of the highly excited state, i.e., when $n\gg 1$, we can approximate $\sqrt{n(n+1)}\approx n\approx\sqrt{n(n-1)}$. Then we get 
\beq
\big\langle E_x(\we{R},t_2)E_x(\we{R},t_1)\big\rangle = 
\eeq{}\tyla
\beq\quad
\frac{e^2}{\big(4\pi\epsilon_0 R\big)^2}\,\omega^4x_0^2\, n \cos\big(\omega(t_2{-}t_1)\big),
\eeq{5}\\[3pt]
just as in the classical case. The average intensity of radiation given by the first correlation function at $t_2=t_1$ is a~constant. The first correlation function for $t_2>t_1$ gives the spectrum of radiation and, in this case, consists of one line only.

The second-order correlation function is more interesting. For $n\gg 1$, we get
\beq 
\hs*{-2mm} \big\langle E_x^2(\we{R},t_2)E_x^2(\we{R},t_1)\big\rangle=
n^2\Big[1+\cos\big(2\omega(t_2{-}t_1)\big)\Big].
\eeq{}\tyla
\beq
\eeq{6}\\[3pt]
The second correlation function oscillates with the frequency $2\omega$. This tells us that the maxima of radiation occur every half period of the electron motion. Thus, the second correlation function can be used to determine the position of the oscillating particle in the vicinity of a~turning point. The high intensity is due to the large acceleration of the oscillating charge and this takes place when the electron is close to one of the turning points. Thus, if high intensity has been detected at $t_1$, then the electron will reach another turning point half the period later, and the intensity will be high once more. Thus, the time dependence of the second correlation function provides information about the motion of the electron. The information is not complete, as the radiation does not distinguish between the two turning points. It is worth noting that the correlation function allows the detection of the particle close to the turning point in spite of the dipole approximation.
\vs*{4pt}

\section{Classical radiation from Kepler orbit\vs*{4pt}}

Before we discuss radiation from the Rydberg states, we will give a~classical description of motion in the Coulomb field~[21]. If the motion is in the $xy$ plane, the coordinates $x$ and $y$ as functions of time are given by 
\beq
x(t)=a\, \big[\cos\big(\xi(t)\big) - \epsilon\big],
\eeq{}\tyla
\beq
y(t)=a\, \sqrt{1-\epsilon^2}\ \sin\big(\xi(t)\big),
\eeq{}\tyl\tylx
\beq
\eeq{7}\\[3pt]
where
\beq 
\omega t+\varphi=\xi(t)-\epsilon \sin\big(\xi(t)\big), 
\eeq{8}\\[3pt]
where $\varphi$ is an~arbitrary phase. The radial variable $r=\sqrt{x^2+y^2}$ can also be expressed as a~function of time
\beq 
r=a\big[1-\epsilon\, \cos\big(\xi(t)\big)\big] .
\eeq{9}\tyla

The parameters $a$, $\omega$, and $\epsilon$ characterize the trajectory. They can be related to energy and angular momentum in the standard way~[21].

We will also need more general trajectories that differ by an~orientation in the plane of the motion described by the phase $\chi$ and by the phase of the motion, $\varphi$. Thus we define
\beq
X(t)=x(t)\cos(\chi)+y(t)\sin(\chi),
\eeq{}\tyla
\beq
Y(t)=-x(t)\sin(\chi)+y(t)\cos(\chi),
\eeq{}\tyl\tylx
\beq
\eeq{10}\\[3pt]
with $\omega t+\varphi=\xi(t)-\epsilon\sin(\xi(t))$.

The classical description of the radiation of a~charge moving along such an~orbit is found to be in complete analogy to the harmonic oscillator case. We will use the dipole approximation since the size of the orbit is much smaller than the characteristic wavelengths of the emitted radiation. The electric field in the far zone is given by
\beq 
\we{E}(\we{R},t)=\frac{1}{R}\, \we{n}\times \big[\we{n}\times\we{a}(t_{ret})\big],
\eeq{11}\\[3pt]
where $\we{a}$ is the acceleration, and $\we{n}=\we{R}/|R|$. Radiation damping is neglected, as in the previous section.

Also, the Fourier decomposition of the trajectory can be found (see~[21]). Here we will give the Fourier decomposition of the $x$ variable
\beq 
x(t)=\sum_k\exp(\ii k(\omega t+\varphi))\, x_k , 
\eeq{12}\\[1pt]
where
\beq 
 x_k=\frac{a}{2 k}\Big[J_{k-1}(k\epsilon)-J_{k+1}(k\epsilon)\Big],\quad k\ne 0.
\eeq{13}\\[3pt]
A similar formula holds for $y(t)$. This will be used in the next section.\pagebreak

\section{Classical limit of matrix elements\vs*{4pt}}

From now on, we will use atomic units.
 
Consider the quantum description of an~atom in a~highly excited energy eigenstate. We label the states by standard quantum numbers: $n$ --- principle quantum number, $l$ --- angular momentum quantum number, and $m$ --- magnetic quantum number. The energy $E_n$ of this state depends on the principal quantum number $n$ as $E_n=-1/(2n^2)$. We will be interested only in states with $m=l$, thus, we will skip the magnetic quantum number to avoid confusion. This means that the wavefunctions considered in this paper are well concentrated in the $xy$ plane, which is perpendicular to the angular momentum. This can be seen from the explicit form of the spherical harmonics function $|Y_{l,l}(\theta,\varphi)|^2\sim\sin^{2l}(\theta)$ that has a~sharp maximum at $\theta=\tfrac{\pi}{2}$ for large $l$. We will, therefore, not consider the wavefunction dependence along the $z$ axis.

The expectation values of the radiated field depend on the matrix elements of the position operator between the quantum states of the atom, i.e., $\langle\psi_{n,l,l}|\,x\,|\psi_{n',l',l'}\rangle$, where $x$ is the coordinate. In spherical coordinates, $x=r\sin(\theta)\cos(\varphi)$, and a~similar expression is valid for the $y$ coordinate, \m{$y=r\sin(\theta)\sin(\varphi)$.} The wavefunctions $\psi_{n,l,l}(r,\theta,\varphi) = R_{n,l}(r)Y_{l,l}(\theta,\varphi)$ are the standard states of the hydrogen atom, with $R_{n,l}(r)$ describing the radial part of the wavefunction and $Y_{l,l}(\theta,\varphi)$ denotes the spherical harmonics. Because of selection rules, these matrix elements are different from zero only if $l'=l\pm 1$.

The radial part of the matrix element of $r^k$ (for any $k$), i.e., 
\beq 
\int\nolimits_0^{\infty}\dd{}r\ r^{2+k} R_{n,l}(r)R_{n',l'}(r) ,
\eeq{14} 
can be found explicitly in terms of special functions~[22]. In fact, the classical limit of this expression, valid for $n\rightarrow\infty$, $l\rightarrow\infty$ with $l/n = \const$, has been found in~[23]. In this limit, (14) approaches the Fourier transform of the classical trajectory $r_{classical}^k$ for the frequency $\omega=(E_n{-}E_{n'})/\hbar$. The classical trajectory $r(t)$ corresponds to the average energy $E=\tfrac{1}{2}(E_n+E_{n'})$ and the eccentricity $\epsilon = \sqrt{1-(l/n)^2}$. Thus, for the matrix element of $r$, we find for $l'=l\pm 1$ that
\beq
\langle n',l',l'|r|n,l,l\rangle \approx a_0\, \frac{n^2}{2  (n{-}n')}
\eeq{}\tylx 
\beq\quad
\times \Big[J_{n-n'+1}\big((n{-}n')\epsilon\big)-J_{n-n'-1}\big((n{-}n')\epsilon\big)\Big],
\eeq{}\tyla \beq
\eeq{15}
where $a_0$ denotes the Bohr radius, and $\epsilon$ corresponds to the eccentricity of the classical orbit with energy and angular momentum equal to the average of the energies of the initial and final state. It should be noted that (15) is analogous to~(5) for the harmonic oscillator, where $\sqrt{n(n+1)}$ is replaced by $n$ for large quantum numbers~$n$.\pagebreak

\begin{figure}
\includegraphics[width=7.6cm]{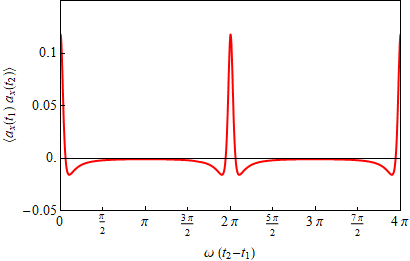}
\caption{The first correlation function of accelerations (23) (normalized to the average square of acceleration) as a~function of $\omega(t_2-t_1)$ for two periods ($\epsilon=0.8$)\vs*{8pt}}\end{figure}

The transition elements for $x$ can also be found 
\beq 
 \big\langle n',l{+}1,l{+}1\big|x|n,l,l\big\rangle+
\big\langle n',l{-}1,l{-}1\big|x|n,l,l\big\rangle  
\eeq{} \tyla\beq 
\qquad \approx x_{n{-}n'},
\eeq{16}\\[2pt]
where $x_{n-n'}$ is given by (13). These formulas allow describing the radiation from the Rydberg states using classical approximations.

The values of the matrix elements can be modeled classically by random trajectories. Consider then the trajectories 
\beq
X(t)=x(t)\cos(\chi)+y(t)\sin(\chi),
\eeq{}\tyla
\beq
Y(t)=-x(t)\sin(\chi)+y(t)\cos(\chi),
\eeq{}\tyl\tyla
\beq\eeq{17}\\[1pt]
with $\omega t+\phi=\xi-\epsilon\sin(\xi)$. The quantities $\phi$ and $\chi$ are random phases, with uniform distributions between $0$ and~$2\pi$. In this case, the expectation values of the $x$ and $y$ operators are equal to the mean values of the classical quantities $X$ and $Y$ with the same values of energy and angular momentum.

\vs*{4pt}
\section{Radiation from a~Rydberg state\vs*{4pt}}

In this and subsequent sections, we will use atomic units in the description of a~quantum state.

The electric field $\we{E}(\we{R},t)$ in the far field is given by the same formula as in the classical field, with the difference that the acceleration $\we{a}$ is an~operator acting on the quantum state of the system consisting of an~electron and the photon vacuum. In the quantum case, also the electric field is an~operator. Thus, for the radiated part of the field, we get in the dipole approximation
\beq 
\we{E}(\we{R},t)=\frac{1}{R}\ \we{n}\times\big[\we{n}\times\we{a}(t_{ret})\big],
\eeq{18}\\[1pt]
where $\we{n}$ is the unit vector in the direction of the observation point, $\we{n}=\tfrac{\we{R}}{|R|}$. 

In what follows, we will find the expectation values of the electric field, as well as the first and second correlation function. It should be noted that the radiation is weak, and therefore the measurement of light intensity in the classical sense is questionable. The expectation value of the electric field squared at a~given point should be understood as the photon counting rate. 

We assume that at $t=0$, the state describes the photon vacuum and the atom is in the state $\psi_{n,l,l}$. This requires matrix elements of the operators $x$ and $y$ and their second derivatives over time.

The first correlation function of the $x$ component of the field radiated in the $y$ direction is given by
\beq 
\langle E_x(\we{R},t_2)E_x(\we{R},t_1)\rangle=\frac{\big\langle a_x(t_{2,ret})\,a_x(t_{1,ret})\big\rangle }{\left(4\pi\epsilon_0 R\right)^2}.
\eeq{}\tyl
\beq
\eeq{19}\\[3pt]
The expectation value of the product of accelerations will be found in the classical limit. First, we will linearize the energy in the vicinity of the initial state energy with the principal quantum number~$n_0$. We get
\beq 
E_n\approx -\frac{1}{2n_0^2}+\frac{n-n_0}{n_0^3}.
\eeq{20} \\[1pt]
\noindent This allows for the approximation of the expectation values of the acceleration operator $a(t)$ by the expectation values of the $r$ operator
\beq
\big\langle n',l-1,l-1\big|a_x(t)\big|n,l,l\big\rangle \approx 
\eeq{}\tyla
\beq\quad
{-}(n{-}n')^2\omega_0^2\, \exp\big({-}\ii(n{-}n')\omega_0t\big)\, x_{n-n'} ,
\eeq{}\tyl\tylx
\beq
\eeq{21}
with $\omega_0=1/{n_0^3}$. Thus, for the two-time correlation function of acceleration in the state $|n,l,l\rangle$, the following can be found 
\beq 
\langle a_x(t_2)a_x(t_1\rangle)=
\sum\limits_{n'l'}\langle n,l,l|x|n'l'l'\rangle\langle n'l'l'|x|n,l,l\rangle
\eeq{}\tylx
\beq\quad
\times \exp\big({\ii \omega (t_2-t_1)(n-n')}\big).
\eeq{22}\\[3pt]
The same can be expressed by the correlation of the classical trajectories
\beq 
\langle a_x(t_2)a_x(t_1)\rangle\approx \int\frac{\dd{} \phi}{2\pi}\int\frac{\dd{}\chi}{2\pi} \frac{\dd{}^2 X(t_2)}{\dd{}t_2^2}\frac{\dd{}^2X(t_1)}{\dd{}t_1^2}.
\eeq{}\tylx
\beq
\eeq{23}\\[3pt]
This is a~good approximation for large $n$ and $l$. The main point is that the matrix elements of the angular part 
\beq 
\hs*{-1mm}\int\hs*{-0.5mm} \dd{}\theta\, \sin(\theta)\,\dd{}\phi\ Y_{l,l}(\theta,\varphi)\sin(\theta)\ee^{\ii\varphi}~ Y_{l-1,l-1}(\theta,\varphi) \nonumber,
\eeq{}\tyla
\beq
\eeq{24}
hence the matrix element of the position operator $x$ weakly depends on $l$ for large~$l$. The correlation function obtained above is shown\linebreak in~Fig.~1.

From the above considerations, it follows that the average intensity of radiation is proportional to the correlation function at $t_1=t_2$ and does not depend on time. The Fourier transform of the correlation function
\beq 
\int \dd{}t\ \big\langle a_x(t)a_x(0)\big\rangle\, \exp(\ii k\omega t)
\eeq{25}\tylx\pagebreak 

\noindent determines the radiation spectrum. Thus the spectrum of radiation from a~Rydberg state can be approximated by the spectrum of radiation from the corresponding classical orbit. 

\vs*{0pt}
\section{Second order correlation\vs*{0pt}}

In this section, we will discuss the second-order correlation function of the radiation originating from a~Rydberg state. This is given by
\beq 
G(t_2,t_1)=\Big\langle E_x(t_1)E_x(t_2)E_x(t_2)E_x(t_1)\Big\rangle.
\eeq{26}\\[3pt]
The state is, as before, the photon vacuum and the Rydberg state of the atom. Expressing the electric field by the acceleration of an~electron in the atom, we get
\beq
G(t_2,t_1)=\frac{1}{R^4}\exp\big({-}2\ii\, n \omega (t_1-t_2)\big)
\eeq{}\tyla
\beq\quad
\times \big\langle a_x(t_1)a_x(t_2)a_x(t_2)a_x(t_1)\big\rangle .
\eeq{27}\\[3pt]
Just as before, we insert a~complete set of states $|n,l,l\rangle$ between the $a$ operators and apply the approximation of $l$ independence of the matrix elements in the case of large $l$. This leads to the following representation of the correlation function\beq
G(t_2,t_1)=\frac{1}{R^4} \int \frac{\dd{}\chi}{2\pi}\int\frac{\rm d\varphi}{2\pi}
\eeq{}\tylx
\beq\quad
\times \frac{\dd{}^2X(t_2)}{\dd{}t_2^2}\frac{\dd{}^2X(t_2)}{\dd{}t_2^2}\frac{\dd{}^2X(t_1)}{\dd{}t_1^2}\frac{\dd{}^2X(t_1)}{\dd{}t_1^2}.
\eeq{28}\\[3pt]
Integration over the angle $\chi$ can be done explicitly, whereas integration over the angle $\varphi$ has to be done numerically.

This is our final result. It gives the second correlation function of radiation emitted by the atom in a~Rydberg state. The formula is approximated and valid for small time differences $t_2-t_1$ because it does not take radiation damping into account. It is valid only in the case of Rydberg states with large $n$ and large $l$, with the maximal magnetic quantum number $m=l$.
 
An example of the second-order correlation function is shown in Fig.~2. One can notice very strong correlation of radiation for small times --- much smaller than the period of motion --- and the periodic behavior of the correlation.

\begin{figure}
\includegraphics[width=7.5cm]{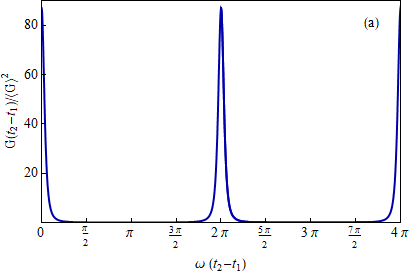}
\includegraphics[width=7.5cm]{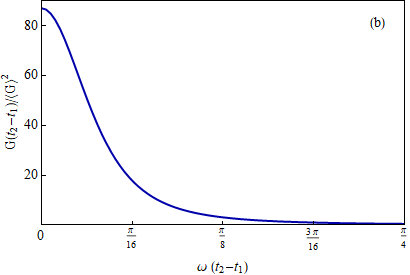}
\caption{The second-order correlation function of accelerations (27) (normalized to the square of the average square of acceleration) as a~function of $\omega(t_2-t_1)$ for one period ($\epsilon=0.8$). Panel (b) shows the same for smaller values of time difference\vs*{0pt}}\end{figure}

\vs*{4pt}\section{Conclusions\vs*{4pt}}

Electromagnetic radiation from an~atom in the Rydberg state can be used to partially localize the electron on the orbit. According to the classical view of radiation, the electron moves along an~elliptic orbit and emits radiation most efficiently when the acceleration is large. This happens when the electron is close to the nucleus. The quantum wavefunction $\psi(r,\theta,\varphi)$ describing the electron state does not indicate the time when the electron is close to the nucleus. Therefore the emitted radiation is time-dependent, and its period reflects the period of motion. The time-averaged intensity, as well as the spectrum of radiation, is constant in time (for a~relatively short time; radiation damping is not taken into account). The second correlation function, $G(t_2,t_1)$, depends on the time difference $t_2-t_1$ and is a~periodic function of time, with the frequency of the classical electron motion. 

In the quantum language, the atom is in a~highly excited Rydberg state with the principal quantum number $n$. The state is stationary, therefore, the average intensity of emitted radiation is constant in time. The spectrum is stationary since radiation damping is neglected, and consists of several narrow lines corresponding to the transition to lower energy states. The second correlation function, however, breaks the time translation symmetry, and this unravels the time evolution of radiation. Based on the measurement of radiation, we can reconstruct the motion of the electron. 

The second correlation function was found in the classical approximation, however, its meaning is indeed purely quantum. The classical approximation means that transition matrix elements have been approximated by the corresponding classical expression. If exact expressions for the matrix elements had been used, the result would have been very similar. The calculations would have been numerically more complex. \pagebreak

We have to stress that electron localization is limited by the uncertainty principle. Thus in the case of the state with orbital quantum number $l$, the angle localization is possible up to $2\pi/l$. While in the case of large $l$ considered here, this is not\linebreak a~strong limitation, it does play a~significant role in the case of $l$ of the order of 1, even for states with large principle quantum numbers $n$.

Our results show that the correlation function is strongly time-dependent. This correlation function clearly shows that if a~strong and short impulse of \m{radiation} is detected, the next such pulse will come after one period of the corresponding classical motion, or in the quantum language, after time $T=2\pi/(E_n-E_{n-1})$. This is due to the large acceleration of an~electron in the vicinity of the nucleus. 

The first strong pulse localizes the electron at this point and breaks the time independence of the radiation. The second pulse comes after one period. Between the strong pulses, the radiation is much weaker because of the small acceleration. Thus the observation of the time dependence of radiation allows the localization of the Rydberg electron in the vicinity of the nucleus.

This method of localizing an~electron on the orbit is non-standard. The recent approach to quantum particle localization is based on successive measurements of a~single particle. Measurement means entangling the particle with another system --- a~pointer --- and then the measurement of the pointer state. In the present approach, the electromagnetic field serves as the pointer. The electron position is not measured directly --- remember the dipole approximation --- the electron acceleration is being measured. Obviously, the second-order correlation gives a~deeper insight into the dynamics than the average values of observables. Also, it provides some insight into the measurement process in quantum mechanics, due to which the difficult process of position measurement is replaced by a~standard measurement of radiation.

We have to point out that the approach described in this paper does not discuss the probabilities of single measurements, but rather it discusses averages such as a~correlation function. Nevertheless, it is a~possible way of detecting the motion of an~electron along a~trajectory.\vs*{4pt}

\AC\vs*{4pt}

JM dedicates this paper to Professor Iwo Bia{\l}ynicki-Birula, who taught me quantum mechanics and for many years guided me through quantum physics.\newpage

\refmake \vs*{4pt}

\re{1}{J. von Neumann, \refdo{\em Mathematical Foundations of Quantum Mechanics}{}, Princeton University, Princeton (NJ) 1955}

\re{2}{W.H. Zurek, \refdo{{\em Rev. Mod. Phys.} {\bf 75}, 715 (2003)}{10.1103/RevModPhys.75.715}}

\re{3}{E.P. Wigner, \refdo{{\em Am. J. Phys.} {\bf 31}, 6 (1963)}{10.1119/1.1969254}}

\re{4}{C.M. Caves, G.J. Milburn, \refdo{{\em Phys. Rev. A} {\bf 36}, 5543 (1987)}{10.1103/PhysRevA.36.5543}}

\re{5}{A.J. Scott, G.J. Milburn, \refdo{{\em Phys. Rev. A} {\bf 63}, 042101 (2001)}{10.1103/PhysRevA.63.042101}}

\re{6}{E.B. Davies, \refdo{{\em Commun. Math. Phys.} {\bf 15}, 277 (1969)}{}}

\re{7}{M.B. Mensky, \refdo{{\em Phys. Rev. D.} {\bf 20}, 384 (1979)}{10.1103/PhysRevD.20.384}} 

\re{8}{A. Barchielli, M. Gregoratti, \refdo{{\em Phil. Trans. R. Soc. A} {\bf 370}, 5364 (2012)}{10.1098/rsta.2011.0515}}

\re{9}{R.J. Glauber, \refdo{{\em Phys. Rev.} {\bf 131}, 2766 (1963)}{10.1103/physrev.131.2766}}

\re{10}{J.R. Klauder, B. Skagerstam, \refdo{\em Coherent States}{}, World Scientific, Singapore 1985}

\re{11}{J. Mostowski, \refdo{{\em Lett. Math Phys.} {\bf 2}, 1 (1977)}{10.1007/BF00420663}}

\re{12}{M. Nauenberg, \refdo{{\em Phys. Rev. A} {\bf 40}, 1133 (1989)}{10.1103/PhysRevA.40.1133}}

\re{13}{A. Rauch, J. Parisi, \refdo{{\em Adv. Stud. Theor. Phys.} {\bf 8}, 889 (2014)}{}}

\re{14}{J.C. Gay, D. Delande, A. Bommier, \refdo{{\em Phys. Rev. A} {\bf39}, 6587 (1989)}{10.1103/PhysRevA.39.6587}}

\re{15}{C.C. Gerry, \refdo{{\em Phys. Rev. A} {\bf 33}, 6 (1986)}{10.1103/PhysRevA.33.6}} 

\re{16}{F. Wilczek, \refdo{{\em Phys. Rev. Lett.} {\bf 109}, 160401 (2012)}{10.1103/PhysRevLett.109.160401}} 

\re{17}{K. Sacha, J. Zakrzewski, \refdo{{\em Rep. Prog. Phys.} {\bf 81}, 016401 (2018)}{10.1088/1361-6633/aa8b38}}

\re{18}{K. Sacha,  \refdo{\em Time Crystals}{10.1007/978-3-030-52523-1}, {\em Springer Series on Atomic, Optical, and Plasma Physics}, Springer, Cham 2020, p.~114}

\re{19}{F. Gampel, M. Gajda, \refdo{{\em Phys. Rev. A} {\bf 107}, 012420 (2023)}{10.1103/PhysRevA.107.012420}}

\re{20}{P. Kowalczyk, C. Radzewicz, J. Mostowski, I.A. Walmsley, \refdo{{\em Phys. Rev. A} {\bf 42}, 5622 (1990)}{10.1103/PhysRevA.42.5622}}

\re{21}{L.D. Landau, E.M. Lifshitz, \refdo{\em Classical Theory of Fields}{}, Pergamon Press, 2000}

\re{22}{L.D. Landau, E.M. Lifshitz, \refdo{\em Quantum Mechanics}{}, Pergamon Press, 2000}

\re{23}{A. Pitak, J. Mostowski, \refdo{{\em Eur. J. Phys.} {\bf 39}, 025402 (2018)}{10.1088/1361-6404/aa997c}}
\end{document}